\RequirePackage{fix-cm}
\documentclass{svjour3}                     
\smartqed  
\usepackage{graphicx}
\begin{document}

\title{Optimising NGAS for the MWA Archive
}


\author{C~Wu         \and
        A~Wicenec \and
        D~Pallot \and A~Checcucci 
}


\institute{C Wu, A. Wicenec, A. Checcucci \at
              International Centre for Radio Astronomy Research\\
              University of Western Australia\\
              \email{[chen.wu, andreas.wicenec, alessio.checcucci]@icrar.org}           
           \and
           D. Pallot \at
              International Centre for Radio Astronomy Research\\
              Curtin University\\
               \email{Dave.Pallot@icrar.org}           
}

\date{Received: date / Accepted: date}

\maketitle

\begin{abstract}
The Murchison Widefield Array (MWA) is a next-generation radio telescope, generating visibility data products continuously at about 400 MB/s. Efficiently managing and archiving this data is a challenge. The MWA Archive consists of dataflows and storage sub-systems distributed across three tiers. At its core is the open source software --- the Next-Generation Archive System (NGAS) --- that was initially developed in ESO. However, to meet the MWA data challenge, we have tailored and optimised NGAS to achieve high-throughput data ingestion, efficient data flow management, multi-tiered data storage and processing-aware data staging.
\end{abstract}

\begin{figure}
  \includegraphics[width=1\textwidth]{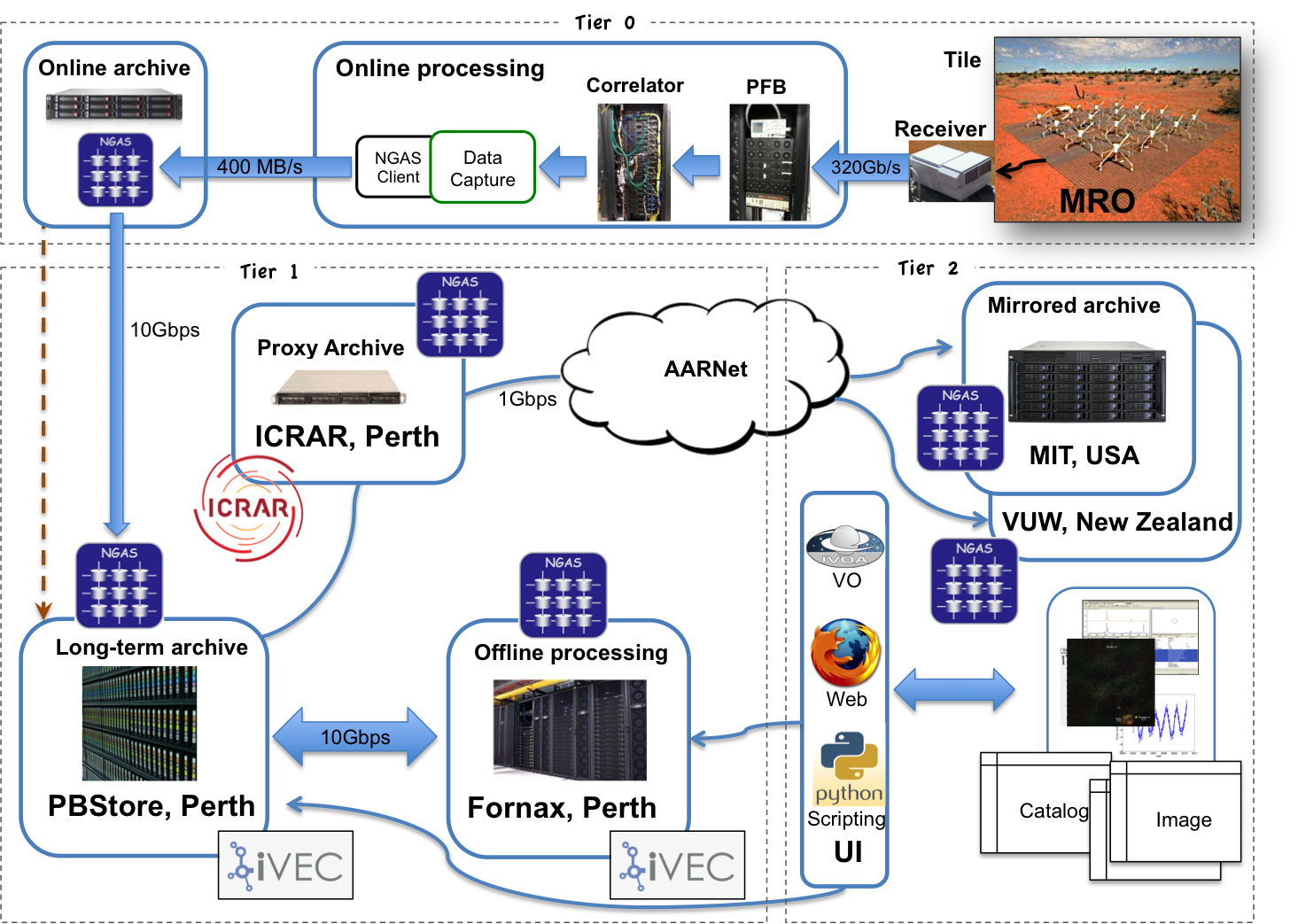}
\caption{The MWA Dataflow consists of three tiers -- Tier 0 is where data is produced, Tier 1 is where data is archived, and Tier 2 is where data is distributed.}
\label{fig:dataflow}       
\end{figure}

\section{Introduction}
\label{sec:1}
The Murchison Widefield Array (MWA) is optimised for the observation of extremely wide fields at frequencies between 80-300 MHz \cite{lonsdale2009murchison}\cite{mwa_overall}. The MWA works without any moving parts, beams are formed by electronically adjusting the phase of the individual dipole antenna, 16 of which form a small phased array known as a ``tile".
The complete array of 128 tiles provides about 2000 $m^2$ of collecting area at 150 MHz. Its instantaneous bandwidth is 30.72 MHz with a spectral resolution of 40 kHz and a time resolution of 0.5 seconds. The field of view of MWA is extraordinary --- between 200 and 2500 square degrees. 

The entire MWA dataflow (Figure \ref{fig:dataflow}) consists of three tiers. Tier 0, co-located with the telescope at the Murchison Radio-telescope Observatory (MRO), includes \emph{Online Processing} and \emph{Online Archive}. Radio-frequency voltage samples collected from each tile (topright corner on Figure \ref{fig:dataflow}) are transmitted to the receiver, which streams digitised signals at an aggregated rate of 320 Gbps to \emph{Online Processing}.  \emph{Online Processing} includes FPGA-enabled Polyphase Filter Bank (PFB) and GPU-enabled software correlators. Correlators output in-memory ``visibilities" that are immediately ingested by the \emph{DataCapture} system. \emph{DataCapture} produces memory-resident data files and uses an \emph{NGAS client} to push files to \emph{Online Archive} managed by the \emph{NGAS} server. 

Each visibility is a complex number representing the amplitude and phase of a signal at a particular point in time within a frequency channel. For a given channel at each time step, the correlator carries out \((N \times (N + 1) / 2) \times (2  \times 2) = 2N(N+1) \) pair-wise cross-correlation and auto-correlation \cite{wayth2009gpu}, where \(N = 128\) is the number of tiles, and \(2 \times 2 \) indicates that polarised samples of orthogonal directions are cross-correlated. 
Since a visibility is a complex number (\(2 \times 32\) bits \(= 8\) Bytes), the aggregated data rate amounts to \( 2N(N+1) \times 8 \times 1/T \times M\) bytes per second, where \(T\) is the time resolution (0.5 seconds) and \(M\) is the total number of frequency channels: \(30.72\) MHz bandwidth \(/\) \(40\) kHz spectral resolution \(= 768\). Hence correlators under the full 128-tile configuration produce visibility data at a rate of 387 MB/s. This requires 24 on-site GPU-based software correlators, each processing 32 channels, and producing frequency-split visibilities at a rate of \(16\) MB/s. Visibilities are currently stored, along with several observation-specific metdata, as FITS image extensions. In addition to visibilities, it is envisaged that images produced by the real-time imaging pipeline \cite{ord2010interferometric} running on those GPU boxes will also be archived online.

At Tier 1 (Perth, a city 700 km south of MRO), \emph{Long-term Archive} (LTA) periodically ingests visibility data stream from \emph{Online Archive} (OA) via a 10Gbps fibre-optic link (i.e. the shaded arrow from OA to LTA), which is a part of the National Broadband Network (NBN) Australia. The dotted arrow in Figure \ref{fig:dataflow} between OA and LTA represents the transfer of metadata on instruments, observations, and monitor \& control information (e.g. telescope temperature). This transfer is implemented as an asynchronous continuous stream supported by the cascading replication streaming environment in Postgresql 9.2. The LTA storage facility --- the Petabyte Data Store (PBStore) --- is a combination of magnetic disks and tape libraries provided by iVEC, a Western Australian organisation with a mission to foster scientific and technological innovation through the provision of supercomputing and eResearch services. From late July 2013, the LTA facility will be gradually migrated to the Pawsey centre \cite{pawsey2013_sgi}, which is a world-class supercomputing centre managed by iVEC with a primary aim to host new supercomputing facilities and expertise to support the Square Kilometre Array (SKA) pathfinder research, geosciences and other high-end science. Both data and metadata at LTA will be selectively transferred to a set of \emph{Mirrored Archives} (MA) at Tier 2. The transfer link between LTA and MAs is optimised by \emph{Proxy Archive} located at the International Centre for Radio Astronomy Research (ICRAR). An additional copy of the data at LTA can be moved to \emph{Offline Processing} for compute jobs such as calibration and imaging. iVEC provides the \emph{Offline Processing} facility --- Fornax, which is a 96-node GPU-based compute cluster located at the University of Western Australia.

Tier 2 currently has two \emph{Mirrored Archives} --- one is at the Massachusetts Institute of Technology (MIT), USA and the other resides in Victoria University of Wellington (VUW), New Zealand. Data in transit from Tier 1 to Tier 2 is carried over by the Australia Academic and Research Network (AARNET) across the Pacific Ocean. These MAs host a subset of the data products originally available in LTA and provide processing capabilities for local scientists to reduce and analyse data relevant to their research projects. While LTA in Tier 1 periodically pushes data to MAs in Tier 2 in an automated fashion, one can schedule ad-hoc data transfer from LTA to a Tier 2 machine via \emph{User Interfaces} (UI). Web interfaces and Python APIs are already available for MWA scientists to either synchronously retrieve or asynchronously receive raw visibility data. We are currently developing the interface compliant to several IVOA standards (e.g. SIAP\cite{ivoa-siap}, TAP\cite{ivoa-tap}) to access MWA science-ready data products such as images (or image cubes) and catalogues.

\section{MWA Archive Requirements and NGAS}
\label{sec:2}
A set of high level requirements of the MWA archive are summarised as follows:
\begin{itemize}
\item High throughput data ingestion (about 400 MB/s) at the telescope site in Western Australia
\item Efficient distribution of large datasets to multiple geographical locations in three continents
\item Secure and cost-effective storage of 8 Terabytes of data collected daily
\item Fast access to science archive for astronomers across Australia, New Zealand, India, and USA
\item Intensive processing (calibration, imaging, etc.) of archived data on GPU clusters
\item Continuous growth of data volumes (3 Petabytes a year), data varieties (visibility, images, catalogues, etc.), and hardware environment (e.g. from iVEC's PBStore to Pawsey supercomputing centre)
\end{itemize}

To fulfil these requirements, we developed the MWA Archive, which is a sustainable software system that captures, stores, and manages data in a scalable fashion. At its core is a Python-based open source software --- the Next-Generation Archive System (NGAS) that was initially developed in the European Southern Observatory (\cite{wicenec2007eso}). The NGAS software is deployed throughout the MWA dataflow, including all archives --- \emph{Online, Proxy, Mirrored, Long-term} --- and \emph{Offline Processing} as shown in Figure \ref{fig:dataflow}. As an open source, NGAS has been previously used in several observatories including the ALMA Archive \cite{wicenec2004alma}\cite{alma_operation}, the NRAO eVLA archive, and the La Silla Paranal Observatory (\cite{wicenec2007eso}).  Distinct features of NGAS include object-based data storage that spans multiple file systems distributed across multiple sites, scalable architecture, hardware-neutral deployment, loss-free operation and data consistency assurance. For example, NGAS manages file metadata, checking data and possibly replicating data to another storage location as fast as possible in order to secure it against loss. Through its flexible plugin framework, NGAS enables in-storage processing that ``moves computation to data".

While NGAS provides a good starting point to develop the MWA Archive, substantial optimisation efforts are required in order to meet MWA data challenges. \emph{First}, the data ingestion rate of MWA telescope at MRO (about 400MB/s) is almost an order of magnitude higher than any existing NGAS deployment with ALMA being the highest --- 66MB/s. It was uncertain whether the existing NGAS architecture scales up to cope with this level of data rate. \emph{Second}, the MWA archive requires efficient flow control and throughput-oriented data transfer to both LTA and MAs. Given the high data volume, the dataflow must thus saturate the bandwidth provided by the public network, and if possible, optimise data routing. \emph{Third}, for cost-effective storage, the archive software needs to integrate ``well" with the commercial Hierarchical Storage Management (HSM) system. For example, the MWA archive needs to interact with the HSM for optimal data staging and replacement based on user access patterns. \emph{Finally}, processing of archived data is essential in MWA to produce science ready data products. Therefore, the MWA archive shall support processing-aware data staging --- moving data from LTA to \emph{Offline Processing} (i.e. the Fornax GPU Cluster) in a way to optimise the parallelism between data movement and data processing. In addition, data management for \emph{Offline Processing} is needed to reduce excessive network I/O contention during data processing.

The remainder of this paper discusses the optimisations we have made on NGAS to meet the above four major challenges. 
\section{NGAS Optimisation and Adaptation}
\subsection{Data ingestion}
\begin{figure}
  \includegraphics[width=1\textwidth]{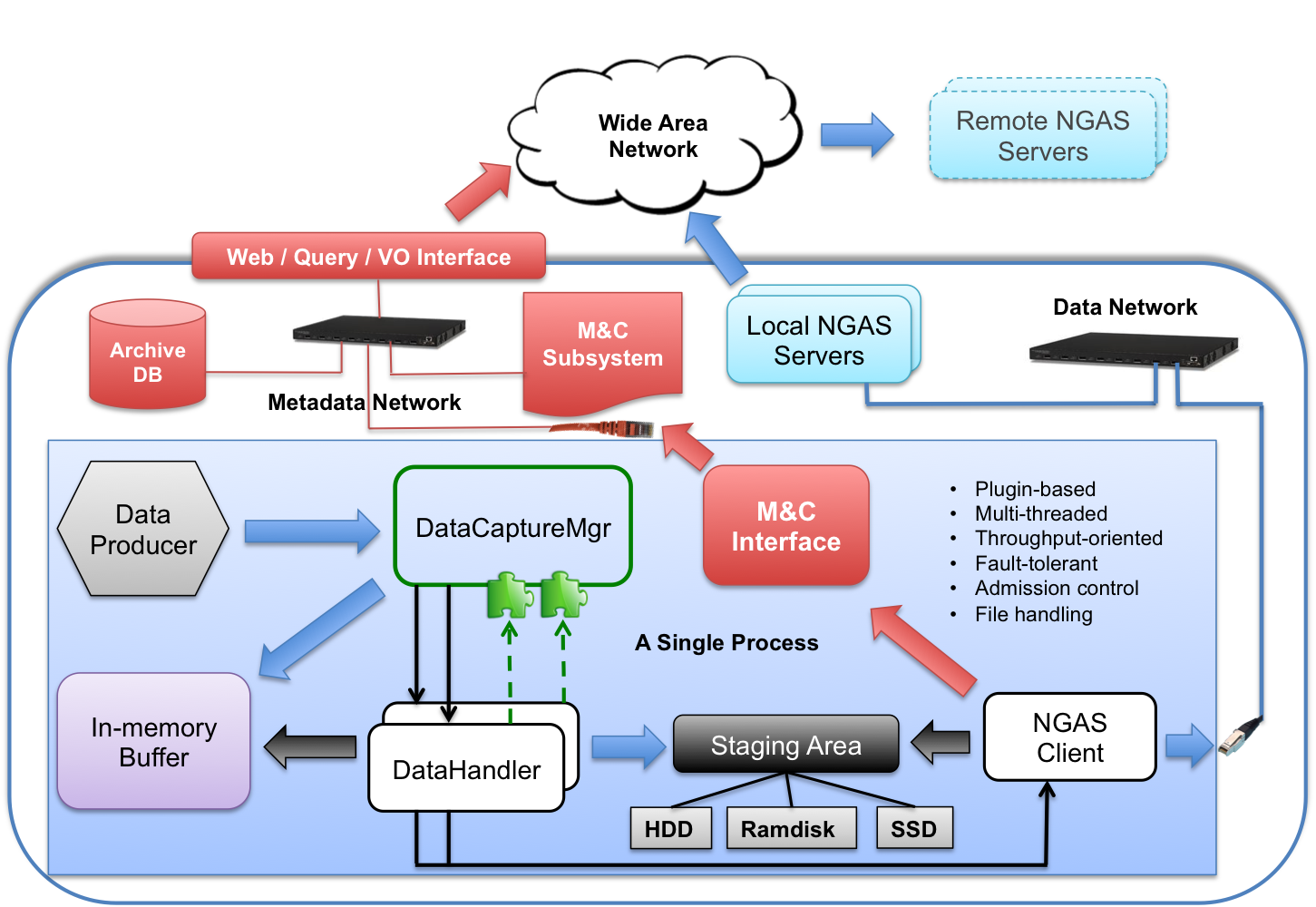}
\caption{Dataflows inside DataCapture --- Blue arrows represent visibility data pushed from one step to the next, dark arrows represent visibilities pulled from the previous step, and red arrows represent the metadata flow. The \emph{NGAS Client} sends file metadata to the \emph{Monitor and Control (M\&C) Interface} and pushes files to the \emph{Online Archive} consisting of two local NGAS servers.}
\label{fig:datacapture}       
\end{figure}

Data ingestion involves fast data capturing and efficient online archiving. Moreover, it requires effective in-memory buffer management and in-process integration with the software correlator. For this purpose, we developed \emph{DataCapture}, which is a throughput-oriented, multi-threaded C++ program with fault-tolerant stream handing and admission control as shown in Figure \ref{fig:datacapture}. 

\emph{DataCapture} follows a plugin-based architecture, where \texttt{DataHandlers} are registered with the \texttt{DataCaptureMgr} to transfer and transform data of various formats (e.g. FITS, HDF5, CASA, etc.). \texttt{DataCaptureMgr} manages a memory buffer, where buckets of memory can be dynamically allocated and deallocated by the \texttt{DataProducer} (i.e. the Correlator) and \texttt{DataHandlers}. The jigsaw puzzle pieces in Figure \ref{fig:datacapture} represent the plugin interfaces by which, data handlers can register themselves with \texttt{DataCaptureMgr}. An important design decision was made that data ingestion is largely pull-based (dark arrows in Figure \ref{fig:datacapture}) via the \texttt{StagingArea}, where feedback throttling control and file handling (e.g filtering, header augmenting, format-aware file chunking, etc.) can be applied asynchronously without interfering with upstream I/O operations.

\emph{Online Archive}  is a cache-mode NGAS optimised for ingesting high-rate data streams. Note that data ingestion is more than just quickly storing data on some disk arrays but also involves metadata management, online check-sum computation, consistency checking, and scheduling data transfer via the Wide Area Network (WAN) to LTA as fast as possible in order to (1) secure it against loss, and (2) vacate more online disks space for buffering. We added a direct triggering mechanism in NGAS so that files can be explicitly scheduled to be removed upon successful delivery while respecting existing cache purging rules (e.g. capacity, age, etc.).  The hardware of \emph{Online Archive} includes two Supermicro nodes, each with 4 quad-core 2.40 GHz Intel Xeon CPUs, 64GB of memory, and a 10Gb Ethernet network host adapter. The storage attached to each node contains 24 Serial-Attached SCSI (SAS) hard disk drives configured as RAID5, providing a total capacity of 80 TB.

\begin{figure}
  \includegraphics[width=1\textwidth]{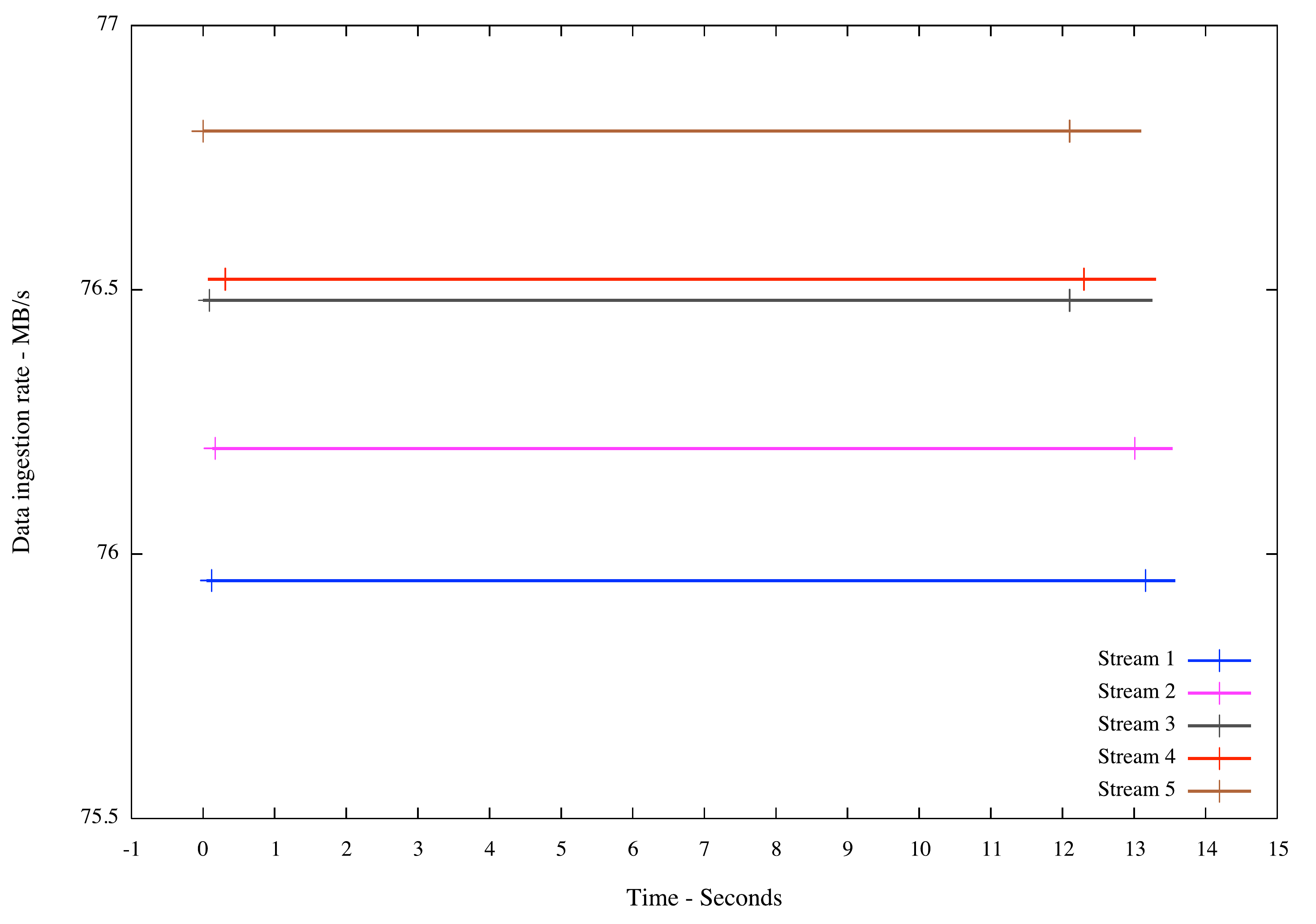}
\caption{Data ingestion at  \emph{Online Archive}. Each one of the five parallel streams records an ingestion rate over 75 MB/s. The aggregated ingestion rate amounts to 382 MB/s}
\label{fig:ingestion}       
\end{figure}

Figure \ref{fig:ingestion} shows the data ingestion rate recorded at \emph{Online Archive} during the early stage of 128-tile commissioning, when five correlator (iDataplex dx360 GPU) nodes are being tested, producing the full data rate (each handles some 153 channels). Each one of the five lines represents a data stream being ingested along the time X-axis (second). The Y-axis measures the sustained ingestion data rate (MB/s) achieved for each data stream.  The fact that all five lines are roughly aligned on both ends suggests that ingestion streams are parallel, yielding an aggregated ingestion rate of around \(76.5 \times 5 = 382\) MB/s. It also shows the variance of data ingestion for a single stream is rather small --- about less than 1 MB/s. Note that this result was achieved on a single NGAS server at  \emph{Online Archive}, which has two NGAS servers. This means each NGAS server only needs to share half of the total ingestion load. It is also worth noting that the recorded mean ingestion rate 76.5 MB/s is not the limit imposed by NGAS, but it corresponds to the maximum rate at which each one of the five correlators can possibly produce data --- about \(387 / 5 = 77.4\) MB/s.

A careful examination of Figure \ref{fig:ingestion} reveals that each line consists of three consecutive sections. The first section \(T_{p}\) records the time spent on I/O preparation --- checking data headers and selecting the appropriate storage medium. The second section \(T_m\), namely the largest section, records the time spent on disk writes and online Cyclic Redundancy Check (CRC) computation. The third section \(T_l\) records the time spent on  post-processing including computing the final destination, quality check, moving data from the staging area to the final destination, metadata registration, and triggering the file delivery event. For all five lines, \(T_{p}\) is almost negligible. Compared to \(T_m\), \(T_l\) is relatively trivial in all five cases, indicating that the post-processing overhead is also very small. This result shows that our data ingestion is optimal --- sequential access  \(T_m\) dominates the I/O.

\subsection{Dataflow management }
\label{sec:dataflow} 
The NGAS software uses a subscription-based mechanism to facilitate dataflows, synchronising data across multiple sites. A data provider has multiple subscribers, each subscribing to data products that will satisfy a subscription condition. Whenever a subscription event is triggered (for example, a new file is ingested into the data provider), a subscriber will receive all data that meet its subscription condition. A subscription condition can be as simple as a timestamp (e.g. ``data ingested since last Friday"). It can also contain complex rules written in an NGAS subscription filter plugin. While NGAS subscription provides a simple and effective method to control the dataflow, it lacks a dedicated data transfer service for bandwidth saturation, network optimisation, etc., which are essential for managing sustained dataflows that move multi-terabytes of data each day across three continents. We therefore made several changes to the existing subscription mechanism. 

First, a single dataflow now supports multiple concurrent data streams, multiplexing the same TCP link to maximise the AARNET bandwidth utilisation. This has significantly improved the cross-Pacific data transfer rate measured from 20 Mbps (a single stream) to 160 Mbps (12 streams), and to 424 Mbps (32 streams), saturating almost half of the maximal bandwidth provisioned by the 1Gbps AARNET. Considering AARNET is a public WAN shared by many users, this transfer rate is reasonably good. In addition to an improvement in the transfer rate, the transfer service is also configurable. For example, the number of streams is adjustable during runtime. Our current research aims to dynamically control this number based on real-time information on network usage.  We have also started the integration of the GridFTP library into the NGAS data transfer module for advanced features such as concurrent transfer of file splits and file pipelining. The data subscribers can be any network endpoint that supports three protocols --- HTTP, FTP, and GridFTP.

Second, we developed a new NGAS command PARCHIVE that enables an NGAS server to act as a \emph{Proxy Archive}, which, as shown in Figure \ref{fig:dataflow}, directly relays a data stream from one network to the next. It works as follows:  a client wishing to transfer a file to destination \(D\) sends the file data via an HTTP POST request (whose URL points to the PARCHIVE command) to the \emph{Proxy Archive}. The  \emph{Proxy Archive} simply reads incoming data blocks into a memory buffer and sends outgoing data blocks from the buffer to \(D\)'s URL, which is a parameter specified in the header of the original HTTP POST request. This is useful when the existing network environment cannot meet the dataflow requirement. For example, the MWA LTA is physically located at PBStore, which provides a storage front-end node called ``Cortex" that is shared by many iVEC users moving lots of data from/to PBStore every day. While iVEC has 10Gb bandwidth for all internal network nodes, Cortex could become the potential bottleneck between iVEC network and the public network. We thus setup a \emph{Proxy Archive} at ICRAR, which sits on the same 10Gb iVEC internal network, but has few shared users moving data to/from the outside world. As a result, MWA data streams bypass Cortex and uses an network endpoint exclusive to MWA to reach remote destinations. Although this relay mechanism can be easily built into the network layer using true proxy servers, placing it in the application layer has an advantage: By writing plugins for the PARCHIVE command, we can perform in-transit processing, transforming (e.g. compressing, sub-setting, aggregating, etc.) data on the move.  This ``processing inside network" approach not only capitalises on the inherent cost of data movement, but potentially reduces storage capacity requirements since no persistent storage is required for in-transit processing.

Third, we created a new NGAS running mode --- \emph{Data Mover}. An NGAS node running in the \emph{Data Mover} mode is a stripped-down version of a regular NGAS server optimised for dataflow management. A \emph{Data Mover} is able to perform all dataflow-related tasks (such as data transfer, subscription management, etc.) on behalf of a set \(A_{normal}\) of normal-mode NGAS servers as long as it has read access to their archived files on storage media. This is achieved by (1) disabling all NGAS commands (e.g. ARCHIVE) that write data into NGAS, which leaves a \emph{Data Mover} completely ``read only" (2) instructing the \emph{Data Mover} to periodically detect changes in NGAS file metadata associated with \(A_{normal}\), and (3) applying subscription conditions to the \emph{Data Mover} against data archived by \(A_{normal}\). In this way, changes in file metadata that meet some subscription conditions will trigger a \emph{Data Mover} to transfer those affected files from current storage media managed by \(A_{normal}\) to external/remote subscribers. This not only removes the burden of dataflow management from \(A_{normal}\), but reduces the coupling between dataflow management and data archiving, separating them in two different processes. An immediate benefit is that, as both archiving and dataflow become increasingly complex, one can change (even shut down) one service without interfering with another, a key requirement for building a scalable and sustainable MWA Archive. Furthermore, we can deploy \emph{Data Movers} as a secure, read-only data management layer that shields internal NGAS nodes from being directly contacted by end users.
\begin{figure}
  \includegraphics[width=1\textwidth]{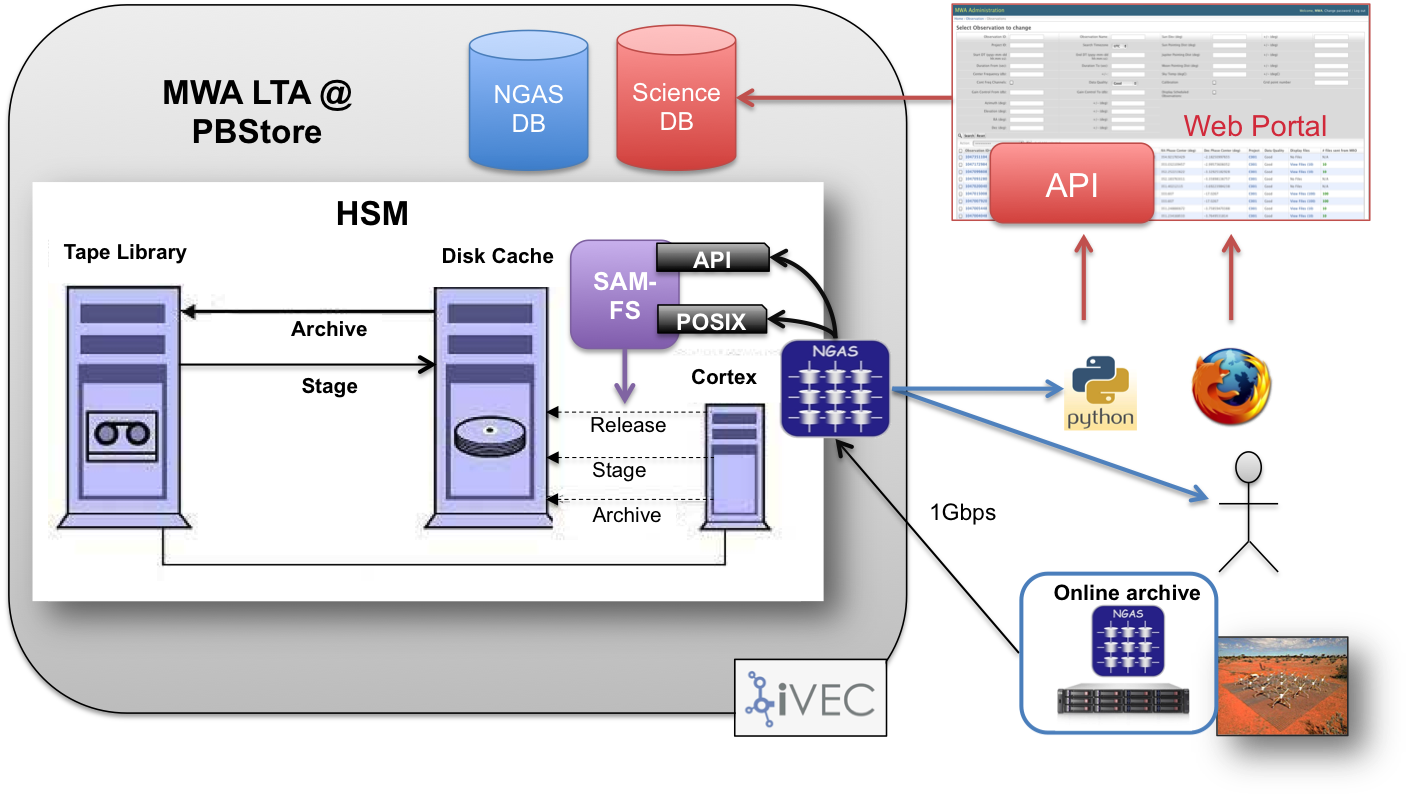}
\caption{Data storage and access at the  \emph{Long-term Archive}}
\label{fig:LTA}       
\end{figure}
\subsection{Data storage and access}
The overall architecture of the MWA \emph{Long-term Archive} is shown in Figure \ref{fig:LTA}. Data on PBStore is managed by the Hierarchical Storage Management (HSM) system --- Oracle SAM-FS, which operates on three levels of storage media: Cortex RAM, disk arrays, and the tape library. Cortex is the front-end machine that has access to data held at PBStore via the POSIX-compliant file system interface provided by SAM-FS.  The \emph{Online Archive} at the telescope site streams data to NGAS archive servers running on Cortex via WAN (10Gb). NGAS servers write this data onto the disk storage using the standard POSIX interface. Throughout the data life cycle, HSM issues data migration requests (\texttt{archive}, \texttt{stage}, and \texttt{release}), moving data across the storage hierarchy. Such movement is either implicitly triggered by POSIX operations (e.g. \texttt{fread()}) or explicitly scheduled based on data migration policies.

While the HSM system automates data migration and abstracts data movement away from application programs, it also provides API libraries that allow applications to directly manipulate data movement between disks and tape libraries. This enables NGAS to optimise data replacement and data staging on LTA.

\paragraph{Server-directed I/O} When applications access data files using POSIX interfaces (\texttt{fread}, \texttt{fseek}, etc.), a missed hit on disk cache will trigger HSM to schedule a \texttt{stage} request that moves data from tapes to disks. This in turn causes HSM to block the current thread for a period time whose duration is non-deterministic in a shared environment like PBStore. In the case of MWA LTA, this might lead to an HTTP timeout or even halt other NGAS threads since not all POSIX operations will gracefully release system resources during I/O wait. To tackle this issue, we added a new NGAS command for asynchronous data access. In the new scheme, an astronomer lists files they wish to access (e.g. all files related to some observations or within a piece of sky) and a URL that can receive these files (e.g. \texttt{ftp://mydomain/fileserver}). After verifying the location of each file, NGAS will simultaneously launch two threads --- a delivery thread that delivers all disk-resident files in the list to this URL, and a staging thread that stages all tape-resident files by explicitly issuing \texttt{stage} requests via the HSM API. The staging thread monitors the staging progress and notifies the delivery thread of any ``new" disk-resident files that have become deliverable. 

Since NGAS now explicitly directs the staging process before performing any I/O tasks on tape-resident files, there are no forced I/O blocking that can affect other NGAS working threads. Moreover, such server-directed I/O  \cite{oldfield2006lightweight}  leaves much more room for optimisation. For example, we let the HSM system reorder the files in the staging list such that files close to one another on the same tape volume will be staged together, reducing many random seeks into fewer sequential reads. This reduction is also possible for different users whose file access lists have a common subset. In addition to server-directed staging, we are investigating server-directed archiving using the declustering technique --- files within a contiguous (temporal, spatial or logical) region will be evenly distributed on multiple tape volumes to enable fast parallel staging.

\begin{figure}
  \includegraphics[width=1\textwidth]{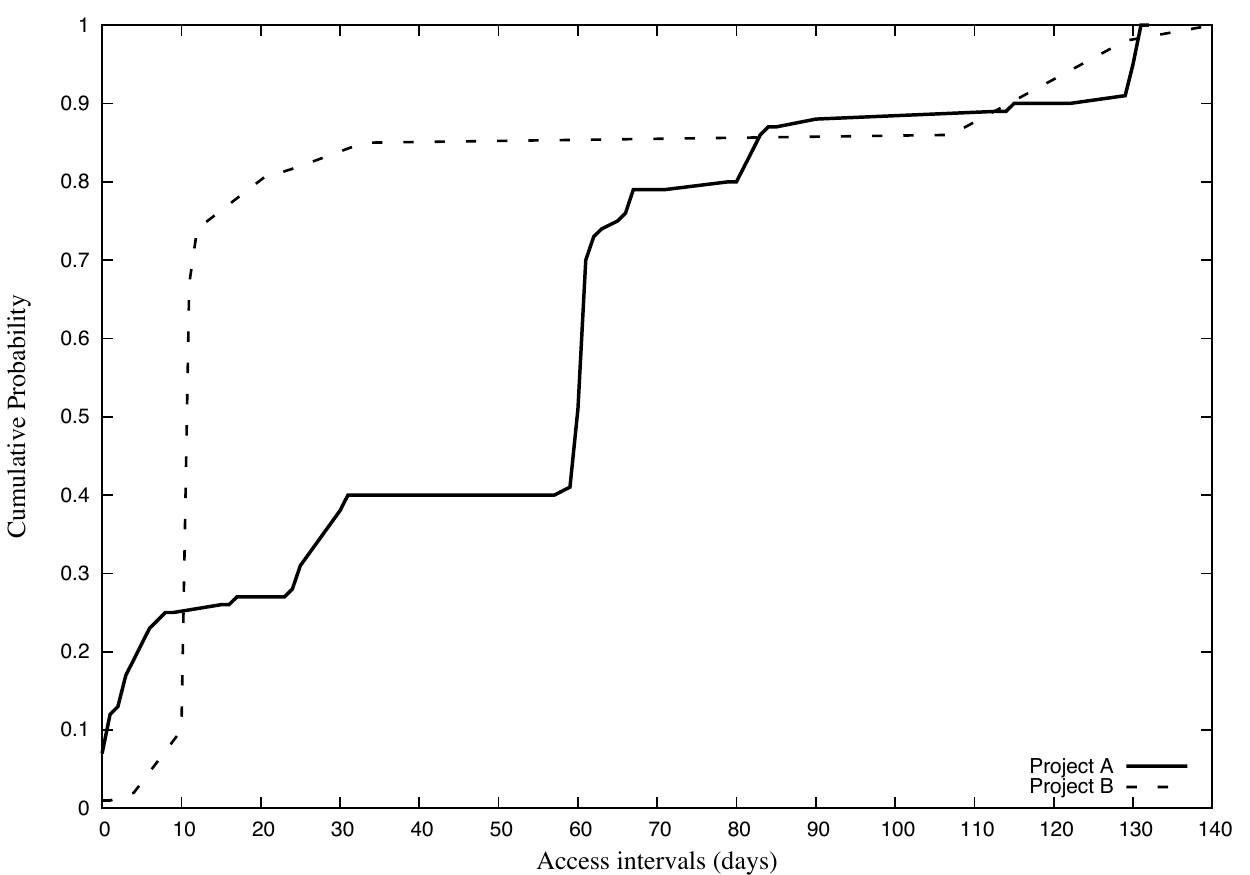}
\caption{The DAPA service produces the empirical distribution of the number of days between inter-accesses recorded at \emph{Long-term Archive} for two science projects during a 140-day period.}
\label{fig:dapa01}       
\end{figure}
\paragraph{Data access pattern analysis}
The analysis of data access patterns becomes essential when the number of missed file hits becomes increasingly large on the HSM disk cache. Although HSM data migration policies based on certain criteria (e.g. high/low watermark, file age/size/type, modified-since-timestamp) can be useful in general, they have several drawbacks. \emph{First}, policies more or less reflect assumptions made by users based on past experience, which could lead to inaccurate assessment on how data is actually being used. An inaccurate parameter in the migration policy could lead to constant system churn, in which, for example, files that have just been released from disks are asked to be staged back from the tape in the next minute. \emph{Second}, data access patterns evolve overtime, and they evolve differently for different types of data and different user groups. Migration policies need to be regularly revisited to reflect such changes, which should be carried out in an automated fashion. \emph{Last}, it is also of a question whether the expressiveness of a predefined migration policy schema can accommodate migration rules based on the ``content" of the data (e.g. FITS header key words) rather than their metadata (such as file size,  age, etc.)

We therefore developed the Data Access Pattern Analysis (DAPA) service base on the existing NGAS logging facility. This service periodically collects raw log files generated by NGAS servers at remote sites, parses the textual description, and converts them into structured records stored in the relational database. DAPA is currently provisioned as a Web service that accepts REST requests to plot various data access statistics. 

Figure \ref{fig:dapa01} shows the empirical distribution of the length of access intervals for two science projects in a 140-day period during MWA commissioning. The X-axis represents the number of days between two consecutive accesses for a file. The two projects appear considerably different, which suggests that content-specific data migration policies can be more optimal. For instance, when the HSM disk cache is running low, we can temporarily release from disks those project B files that have not been accessed for a month. On the other hand, we could pre-stage to disks those tape-resident project A files that have not been accessed for two months given the sharp increase in inter-access probability between 60 days to 90 days. 

Note that we cannot predict precisely how data will be accessed in the future since the true process governing astronomers to access certain data at a given time is overly complex and hence cannot be modelled as such. Our approach assumes that, when more scientists access the same data archive for an extended period of time, the overall process can be considered as random and hence can be simplified using probability models.  The development of DAPA service is a proof-of-concept of this approach and is still at an early stage. In the next step, we will fit various probability models based on these empirical distributions. Ultimately, these fitted models and other useful principles can establish objective functions (e.g. minimise disk cache usage, expected access latency, miss rate, etc.), re-casting data migration as an optimisation problem.

\subsection{Data staging \& processing}
This section discusses data management for running MWA data processing pipelines on a GPU cluster - Fornax \cite{harris2012gpu}. Fornax is a data-intensive SGI supercomputer that comprises 96 compute nodes, each containing two 6-core Intel Xeon X5650 CPUs with 72GB RAM and a NVIDIA Tesla C2075 GPU with 6 GB RAM. To the MWA Archive, the distinct ``data-intensive" feature of Fornax are: (1) In addition to the traditional ``global" storage (500 TB managed by the Lustre filesystem), each compute node has a directly-attached, RAID-0 disk array with a capacity of 7 TB, which amounts to a total of 672 TB ``local" storage. (2) two InfiniBand networks (40Gbps) with the first one used for the Message Passing Interface (MPI) interconnect and the global storage network and the second dedicated to data movement across local disks on neighbouring compute nodes, (3) two Copy Nodes each has a dedicated 10Gb network link to the outside world. 
\begin{figure}
  \includegraphics[width=1\textwidth]{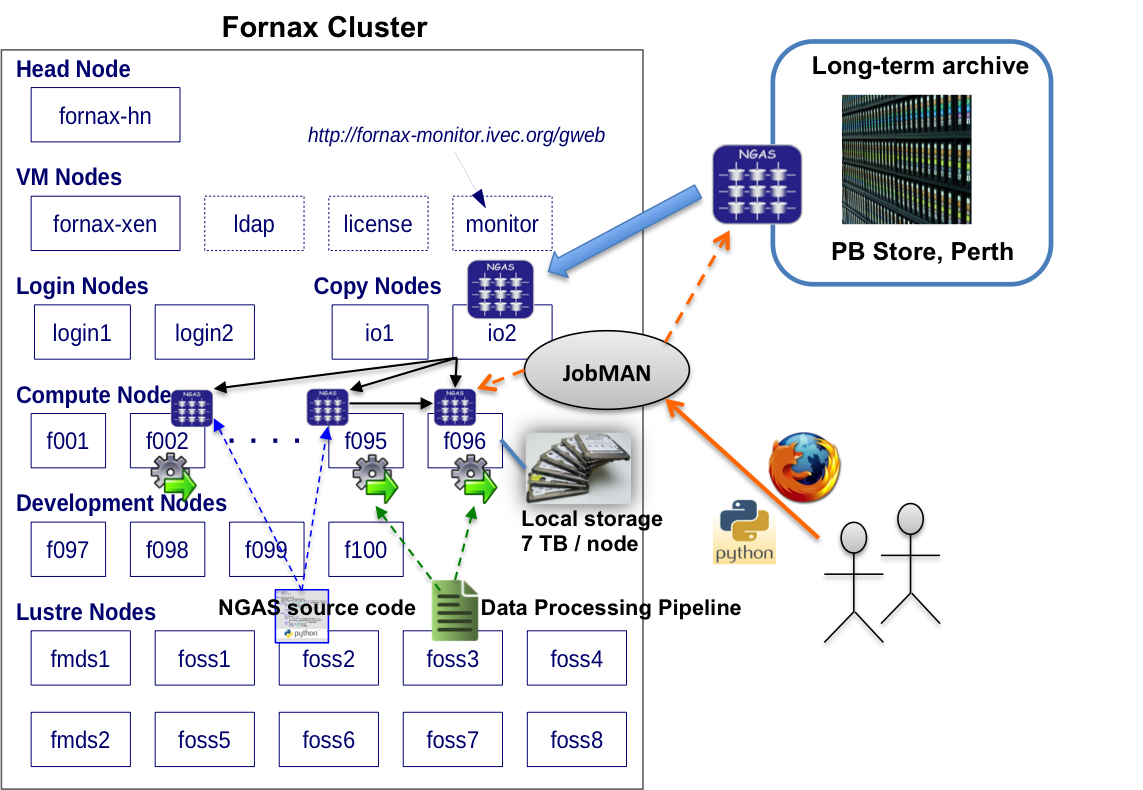}
\caption{Data staging and processing at \emph{Fornax Cluster}, which includes 96 Compute nodes, 2 Copy nodes, 6 Lustre storage nodes, 2 Lustre metadata server nodes, and so on. Each compute node has 7 TB of local storage. Both NGAS code and the Data processing pipeline are kept on the storage nodes and are accessible and executable from within a compute node via a shared file system --- Lustre. Initially, NGAS servers are launched by the cluster job scheduler on compute nodes, which are dynamically allocated for processing NGAS jobs. Next, scientists (bottom right) issue data processing requests to the  \emph{JobMAN} Web service via browsers or Python libraries.  \emph{JobMAN} analyses the request, queries related NGAS servers on compute nodes. For data only available at LTA,  \emph{JobMAN} sends a data staging request to LTA NGAS servers. LTA arranges data migration from tape libraries, and streams data to the NGAS server running on a Copy node, which uses the PARCHIVE command to forward the data stream to a compute node where it will be processed by an NGAS processing plugin. For data already kept on Fornax, \emph{JobMAN} schedules data movement directly from one NGAS server to another where processing will take place. Once all data required by a task in a processing pipeline is available on a compute node, NGAS server executes (gears with green arrows) that task after loading the executable into the compute node and returns the results back to \emph{JobMAN} for scientists to retrieve.}
\label{fig:staging}       
\end{figure}

Two important goals guided our design: (1) to create overlaps between data staging and data processing, and (2) to utilise the 7 TB local storage attached to each compute node. The first goal aims to achieve a higher level of parallelism between I/O and computation. Nevertheless this is not straightforward on most of today's supercomputing clusters (including Fornax), where scientists often have to launch two separate jobs --- one for data staging, and the other for processing. The processing job typically waits for the staging job to finish copying all required data from remote archives to the cluster. Introducing overlaps between these two jobs allows some processing to be done while data is being staged, thus reducing the total completion time by ``hiding" I/O behind computation.

The second goal aims to achieve satisfactory I/O performance with a low variability during data staging. Unlike traditional compute-intensive clusters where all user data is centrally stored and managed by a global storage network consisting of several dedicated storage nodes, data-intensive clusters (e.g. Fornax and DataScope \cite{data_scope}) provide compute-attached disk arrays. Since writing data to local disks does not involve any network I/O --- which is somewhat volatile considering many applications/users are sharing the cluster resource at a given time, applications using local storage can thus achieve good I/O performance with a smaller variance. In fact, the aggregated I/O throughput across all 96 compute nodes is in the order of 40 GB/s, far beyond what a centralised shared file system can normally reach. Furthermore, as local disk arrays are evenly distributed across the entire cluster, they have a better load balancing, scalability, and are less susceptible to a single-point-failure. However, traditional cluster job schedulers rely solely on a shared file system to manage the global storage, but lack the capability to provide applications with access to node-attached local disks. As a result, these high performance local storage resources on Fornax are often under-utilised.

Figure \ref{fig:staging} depicts the overall solution to achieve the above two goals. At its core is the NGAS Job Manager (\emph{JobMAN}), a Web service that schedules job executions to NGAS servers on Fornax compute nodes. To interleave processing with data staging, \emph{JobMAN} decomposes a processing request into small tasks that can be independently executed in parallel. Such decomposition is domain specific. In the case of MWA imaging/calibration, visibility data in each observation is split into a number of sub-bands, each containing a fixed number of frequency channels. Data in different sub-bands are staged and processed in different tasks on given compute nodes. To select an ideal compute node for processing a task, \emph{JobMAN} queries the Fornax NGAS database to obtain data locality information on local storage, and triggers necessary data movement (e.g. from LTA to a compute node, or from one compute node to another) via the NGAS subscription and asynchronous retrieval features discussed in Section \ref{sec:dataflow}.

To utilise the local storage, we essentially used a two-level (cluster job scheduler and \emph{JobMAN} scheduler) scheduling technique \cite{Raicu2008}. In the first level, NGAS servers acquire compute node resources through the normal cluster job scheduler on Fornax --- PBS (Portable Batch System) Pro. Once resources are acquired, in the second level, \emph{JobMAN} is in charge of matching tasks with ``correct" data on some compute nodes, and moves data if necessary using NGAS dataflow management system (Section \ref{sec:dataflow}). In this way, each task is paired with required data at its local disks that cannot be utilised by a shared file system or traditional job schedulers (e.g. PBS Pro). 

In summary, NGAS \emph{JobMAN} allows for parallel executions of staging and processing, reducing the overall time-to-value for data products. It also reduces the network I/O contention by restricting each task to access local storage resource. We are currently working on numerous interesting topics in this line of work including the avoidance of hotspots during NGAS job scheduling that requires partial remote I/O access, optimal staging strategies \cite{Bharathi2009}, and dealing with iterative workflows with data-dependency and intermediate data products.

\section{Conclusion}
As a precursor project to the SKA-Low \cite{skalow-mwa}, the MWA is a next-generation radio telescope that generates large volumes of data, opening exciting new science opportunities \cite{bowman2013science}. The MWA Archive has stringent requirements on data ingestion, transferring, storage \& access, staging and processing. NGAS is a feature-rich, flexible and effective open source archive management software that has been used for several large telescopes in both the optical \cite{wicenec2007eso} and radio \cite{alma_operation} astronomy community. However, to tackle the MWA data challenge, we need to tailor and optimise NGAS to meet all these requirements. In this paper, we discussed these optimisations and evaluated methods we have used to achieve high-throughput data ingestion, efficient data flow management, multi-tiered data storage and optimal processing-aware staging. 

Thus far, the MWA Archive has met the requirements during the commissioning phase. For example, \emph{Online Archive} is able to cope with the full ingestion data rate. The MWA dataflow has synchronised data across three tiers and saturated available network bandwidth to maximise the data distribution efficiency. The turnaround time --- from observing an area of the sky to data products appearing in the Tier-2 hosts --- can be reduced to minutes. This is achieved by moving about 3 {\raise.17ex\hbox{$\scriptstyle\sim$}} 4 TB of data from Western Australia to MIT, USA every day. The MWA LTA is currently being relocated to the new iVEC Pawsey supercomputing centre \cite{pawsey2013_sgi} for the full operation phase.

The entire MWA Archive software environment is rather generic --- the Python-based NGAS software currently runs on both Linux-based (CentOS, SUSE Linux Enterprise, and Ubuntu Server) and UNIX-based (Solaris) operating systems. The NGAS software requires several standard third-party software libraries such as \texttt{python2.7-devel}, \texttt{autoconf}, \texttt{zlib}, \texttt{readline}, \texttt{openssl} and so on, all of which are installed during the automated installation process using one of the standard distribution package managers such as \texttt{RPM}, \texttt{YUM}, \texttt{Zypper} or \texttt{APT}. The NGAS software has adopted a highly flexible, plugin-based software architecture where the core does not make any special assumptions about the hardware, operating systems or file systems. As a result, all computing environment-specific optimisations made for the MWA Archive (e.g. querying file location status on tape libraries, interaction with the HSM) are provided as NGAS plugins, which can be dynamically plugged into and removed from the NGAS core based on the run-time configuration.

During a period of sixteen months, four authors of this paper collectively contributed one Full-Time Equivalent (FTE) time towards developing the MMA Archive --- from hardware procurement \& configuration, network setup \& tuning to software implementation \& test, database configuration, and software deployment. The code base of the NGAS core software contains about 40,000 lines of Python code. For the MWA Archive, we have added about 5,500 lines of Python code, most of which in the form of NGAS plugins and 4,000 lines of C++ code to develop the \emph{DataCapture} sub-system. The NGAS software has a Lesser GPL license and can be obtained upon request. We are currently in the process of forming the NGAS user group, inviting members from around the world who are using NGAS as their archive management software. We envisage that the NGAS user group will release the software to the public domain in the future. 

\section{Acknowledgment}       
The authors would like to thank iVEC for providing the PB Store, Fornax and Pawsey supercomputing facilities. The Pawsey centre is funded from Australian federal and Western Australian state grants. ICRAR is a joint venture between the Curtin University, the University of Western Australia and received grants from the Western Australian Government. The authors would like to acknowledge the excellent work of the original main developer of NGAS, Jens Knudstrup from ESO, without which this work would have been impossible. The authors would also like to thank two anonymous reviewers for providing valuable feedback to the original manuscript.

%


\bibliographystyle{spphys}       


\end{document}